\documentclass[twocolumn,showpacs,preprintnumbers,amsmath,amssymb]{revtex4-1}

\usepackage{epsfig}
\usepackage{lipsum}
\usepackage{tabularx}
\usepackage{array}
\usepackage{mathrsfs}
\usepackage{amssymb}
\usepackage{amsmath}
\usepackage{cases}  
\usepackage{graphicx}
\usepackage{subfigure} 
\usepackage{dcolumn}
\usepackage{bm}
\usepackage{verbatim} 
\usepackage[dvipdfm,pdfstartview=FitH,colorlinks,linkcolor=blue,anchorcolor=blue,citecolor=blue]{hyperref}
\usepackage{amsthm}  

\begin{document}

\title{Free-space continuous-variable quantum key distribution of unidimensional Gaussian modulation using polarized coherent-states in urban environment}

\author{Shi-yang Shen}
\author{Ming-wei Dai}
\author{Xue-tao Zheng}
\author{Qi-yao Sun}
\author{Bing Zhu}
\author{Guang-Can Guo}
\author{Zheng-Fu Han}
\affiliation{Key Laboratory of Quantum Information, University of Science and Technology of China, Hefei 230026, China}


\begin{abstract}

We use single homodyne detector to accomplish Continuous-Variable quantum key distribution(CV QKD) in a laboratory and urban environment free-space channel. This is based on Gaussian modulation with coherent-states in the polarization degree of freedom. We achieved a QKD distance at 460m, at the repetition rate of 10 kHz. We give the security of this protocol against collective attack in the asymptotic regime. The secure key rate is 0.152 kbps at the typical reconciliation efficiency of 0.95. The experiment setup of this scheme is simplified and the difficulty to realize has been remarkably reduced compared to traditional symmetric modulation ones, for example, GG02 protocol. The influence of security key rate brought by asymmetric modulation is small in a relative low channel loss condition in the free-space environment. This scheme is expected to be significance meaning to the future practically utilize.

\end{abstract}

\pacs{}

\keywords{}

\maketitle

\section{introduction}
\label{intro.}

Quantum key distribution allows the two authorized distant parties, Alice and Bob,  to share a common key via a potential eavesdropped quantum channel. The first QKD protocol was been proposed in 1984 \cite{BB84}. While CV QKD protocols, especially ones with coherent-states light source, has been concerned recently \cite{GG02, reverse rec, reverse 2}, which utilize balanced homodyne detection technique, and light source is not at single photon level. Therefore, it has advantages of higher detection efficiency, thus higher secure key rate, and anti photon number attack. During a few years of development, CV QKD protocols and experiment implementations have been modified and simplified. First, the coherent-states protocols show substantial advantage against squeezed-states versions\cite{squeezed prot, squeezed, squeeze polar, squeeze free space} in the prepare of light source, and theoretical secure distance improves remarkably, which leads to the deep research of CV QKD theory and realizing variable experiment schemes. As far as the process of present experiments in the field of CV QKD, the symmetrically Gaussian modulated coherent-state(GMCS) protocols has been quite well studied\cite{GG02 experiment, free space exp, free space feasible, np exp, long distance, fiber exp} since the composable security analysis has been revealed \cite{composable}. Secondly, instead of Gaussian modulation, the discrete modulation reduces the complexity in the classical post-processing, in which case the signal-to-noise ratio is low because of the long distance propagation loss. Finally, the unidimensional CV QKD protocol proposed in 2015\cite{unidimension} further simplified apparatus in both preparing and detection sides since only one of the quadratures should be Gaussian modulated instead of both being modulated simultaneously. Experiment scheme in fiber channel has been accomplished in \cite{unidimension exp} and the security key rate at finite size scenario has been proved\cite{ud finite key}.  Furthermore, composable security of unidimensional CV QKD has been revealed in \cite{ud composable}.

Besides, free-space channel is insensitivity to polarization compared to fiber channel, which results to be unchanged of the light polarization as propagating. Thus the polarization controller at receiver's side can be left out. In another words, the system needn't calibrate polarization direction frequently, reducing the calibrating time of non-key distribution, and thus increase the key rate. On the other hand, encoding with polarization avoids the non-synchronous disturbance of the phase. Therefore, the phase locking between local oscillator and signal is unnecessary as soon as the polarization has been aligned to the same direction, significantly simplifying the difficulty of system implementation. Security distance of CV QKD in free-space\cite{free space exp, free space feasible, squeeze free space} channel can reach dozens of kilometers, compatible with urban condition communication. It is expected to play an important role in the future practical applications.

This experiment uses unidimensional CV QKD scheme in the free-space channel, modulating the polarization quantum Stokes parameter with Gaussian distribution and gets the security key rate at real urban environment condition of 460m, which turned out to be little less performance but obviously simplified and more adaptable to experimental environment, compared to GG02 protocol at the same conditions.

The article is organized as follows, in section II and III, we explain the principle of coherent-state encoded in polarization degree and give the security key rate analysis against collective attack. In section IV, the experimental setup and environment will be revealed. Finally, in section V, we deal with the raw key data and give the security key rate.

\section{protocol description}

In section II, we describe the prepare and measure version of unidimensional protocol proposed in \cite{unidimension}, which correspond to the real experiment system implementation. In usual CV QKD protocols, both quadratures, X and P, must be modulated simultaneously. However, in this protocol, the situation is contrast. Only single quadrature, without loss of generality, denoted as X, will be modulated. Each coherent-state, is displaced by x in phase space, which obeys Gaussian distribution centered at 0 and has the variance of $V_{M}$. And the other quadrature has the variance of 1,normalized at the shot noise unit(SNU). Bob performs homodyne detective X quadrature in a certain time interval and sometimes monitor variance of P quadrature as a channel parameter. After Alice and Bob share a sufficient long sequence of real number raw key data, they estimate the channel parameters using a small random part of the data and perform reverse-reconciliation \cite{reverse rec}.

\label{Polarization encode principle}
For polarization encoding, quantum Stokes operators are treated as quadratures like phase encoding scheme, defined as follow \cite{stokes op}:
\begin{equation}
\begin{aligned}
\hat{S}_{0}&=\hat{a}^{\dagger}_{H}\hat{a}_{H}+\hat{a}^{\dagger}_{V}\hat{a}_{V},\\
\hat{S}_{1}&=\hat{a}^{\dagger}_{H}\hat{a}_{H}-\hat{a}^{\dagger}_{V}\hat{a}_{V},\\
\hat{S}_{2}&=\hat{a}^{\dagger}_{H}\hat{a}_{V}+\hat{a}^{\dagger}_{V}\hat{a}_{H},\\
\hat{S}_{3}&=i(\hat{a}^{\dagger}_{V}\hat{a}_{H}-\hat{a}^{\dagger}_{H}\hat{a}_{V}),\\
\end{aligned}
\end{equation}
where subscripts H and V label the creation and annihilation operators along horizontal and vertical polarization mode, respectively. These creation and annihilation operators obey the same commutation relations and Heisenberg uncertainty principle as the quadratures X and P in phase space except a constant coefficient:
\begin{equation}
[\hat{a}_{j},\hat{a}^{\dagger}_{k}]=\delta_{jk}, j,k=x,y
\end{equation}
and
\begin{equation}
[\hat{S}_{j},\hat{S}_{k}=2i\epsilon_{l}\hat{S}_{jkl}], for j,k,l=1,2,3
\end{equation}
While the variance of the latter three Stokes operators satisfy
\begin{equation}
\label{uncertainty}
Var[\hat{S}_{2}] Var[\hat{S}_{3}]\ge |\langle\hat{S}_{1}\rangle|^{2}
\end{equation}
In our work, we use polarization degree to encode information. The $S_{1}$ polarized (V mode) coherent-state light plays the role of local oscillator(LO). $S_{3}$ and $S_{2}$ polarized state generated by a electro-optical modulation(EOM) are two orthogonal quadratures. Since the intensity of the modulated light is far weaker (about 3 orders of magnitude lower) than the LO, the loss of LO is negligible, and the intensity of circle polarized light $|S_{1}|$ nearly remain unchanged. In other words, the right hand of equation (\ref{uncertainty}) is approximatively a constant. The output light is at strong vertical polarized mode with a superposition of a very weak circle mode. More explicitly, the shape of the polarization state in x-y space is an ellipse with a eccentricity of nearly unity, and its long and short axis are oriented to the direction of modulation intensity.

If the applied voltage of EOM is $U$, the phase difference between ordinary and extraordinary light is $\phi=\pi U/V_{\pi}$,  where $V_{\pi}$ is the half wave voltage of EOM. Assume that the annihilation operator of input light is:
\begin{equation}
\hat{a}_{in}=\frac{a_{LO}}{\sqrt{2}}\left[
\begin{array}{ccc}
0\\
1\\
\end{array}\right]
\end{equation}
After EOM, without loss of generality, except for a phase factor, the output light is
\begin{equation}
\hat{a}_{in}=\frac{a_{LO}}{\sqrt{2}}\left[
\begin{array}{ccc}
0\\
e^{i \phi}\\
\end{array}\right]
\end{equation}
The light then passes through a quarter wave plate whose fast and slow axises is $45\deg$ to the axises of EOM, and the Jones Matrix of QWP is
\begin{equation}
J_{QWP}=\left[
\begin{array}{ccc}
\cos\frac{\pi}{4}&-\sin\frac{\pi}{4}\\
\sin\frac{\pi}{4}&\cos\frac{\pi}{4}\\
\end{array}\right]
\left[
\begin{array}{ccc}
1&0\\
0&i\\
\end{array}\right]
\left[
\begin{array}{ccc}
\cos\frac{\pi}{4}&\sin\frac{\pi}{4}\\
-\sin\frac{\pi}{4}&\cos\frac{\pi}{4}\\
\end{array}\right]
\end{equation}
Then the light is
\begin{equation}
\hat{a}=J\cdot \hat{a}_{out}=
\frac{a_{LO}}{2}\left[
\begin{array}{ccc}
1+i e^{i \phi}\\
1-i e^{i \phi}\\
\end{array}\right]
\end{equation}
After a 50:50 PBS and a balanced homodyne detector, the measured photon number difference is
\begin{equation}
n_{meas}=\hat{a}^{\dagger}_{H}\hat{a}_{H}-\hat{a}^{\dagger}_{V}\hat{a}_{V}=a^{2}_{LO}\sin{\frac{\pi U}{V_{\pi}}}
\end{equation}

When $U\ll V_{\pi}$,
\begin{equation}
\label{var}
n\approx a^{2}_{LO}\frac{\pi U}{V_{\pi}}
\end{equation}
Thus, if $U$ obeys Gaussian distribution, $U\sim N(0, \Sigma^{2})$, then $n\sim N(0,a^{4}_{LO}\frac{\pi^{2}\Sigma^{2}}{V^{2}_{\pi}})$.
The relation between applied voltage on EOM and modulation variance $V_{M}$ is
\begin{equation}
V_{M}=\frac{\pi^{2}\Sigma^{2}V^{2}_{LO}}{V^{2}_{\pi}N_{0}}
\end{equation}
according to Eq. \ref{var}, where $V_{LO}$ is the voltage of local oscillator in Alice's side measured by DAQ. However, the quadrature's variance is in SNU, the Stokes operator must be normalized as X and P. On one hand, the expectation value
\begin{equation}
\begin{aligned}
\langle\hat{S}_{3}\rangle&=\langle a_{H}a_{V}|i(\hat{a}^{\dagger}_{V}\hat{a}_{H}-\hat{a}^{\dagger}_{H}\hat{a}_{V})|a_{H}a_{V}\rangle\\
&=2a^{2}_{LO}\sin\phi\\
&=2n_{meas}\\
\end{aligned}
\end{equation}
On the other hand, note that the local oscillator is strong enough so that $\langle S_{1}\rangle\approx a^{2}_{LO}$, donated as $S_{1}$. By defined two new operators
\begin{equation}
\begin{aligned}
\hat{X}&=\frac{\hat{S}_{2}}{\sqrt{S_{1}}}\\
\hat{P}&=\frac{\hat{S}_{3}}{\sqrt{S_{1}}}\\
\end{aligned}
\end{equation}
$\hat{X}$ and $\hat{P}$ are evidently have the same form of commutation relations and Heisenberg uncertainty principle as usual defined quadrature X and P. To be distinguishable, we still use new symbols X and P instead of $S_{2}$ and $S_{3}$ below, unless either mentioned. Therefore, $X_{a}$ and $X_{b}$ are proportional to average photon numbers, thus proportional to the voltages measured by DAQs since the photo-diodes work on linear mode.
\begin{equation}
\begin{aligned}
X_{a}&=\frac{2n_{prep}}{\sqrt{n_{a}}}\\
X_{b}&=\frac{2n_{meas}}{\sqrt{n_{b}}}\\
\end{aligned}
\end{equation}
where $n_{a}=a^{2}_{LO}$ and $n_{b}=T \eta n_{a}$ are the average photon number of local oscillator of Alice and Bob respectively, and they can be monitored by power meter or wave oscilloscope in the unit as light intensity or voltage. And $T, \eta$ are overall transmittance and detection efficiency of homodyne.

\section{security analysis}

The security of CV QKD has been studied especially in recent years. \cite{gaussian extrem} give the extremality of Gaussian states, and as a consequence, the extremality of Gaussian attacks in \cite{gaussian attack extrem 1, gaussian attack extrem 2} against collective attack scheme in the asymptotic region. In 2010\cite{finite}, the finite-size effect analysis of CV QKD has been shown also by Leverrier. The security against general attacks in practical finite-size region has been proven in 2013 \cite{general attack}, which exploits symmetry in phase-space even with post-selection, as far as the modulation and post-election is symmetric in phase-space. Soon later, Leverrier\cite{composable} achieved the composable security for coherent CV QKD protocols against collective attacks, which established the security of coherent protocols against general attacks.

The security key rate for the unidimensional protocol is computed in \cite{unidimension, ud finite key} against collective attacks in asymptotic and finite size region. As already obviously known, the lower bound key rate is given by
\begin{equation}
K=I_{AB}-\chi_{BE}
\end{equation}
where
\begin{equation}
\chi_{BE}=S(E)-S(E|x_{B})
\end{equation}
is the Holevo information \cite{holevo, unidimension} between Bob and Eavesdropper in the scheme of reverse reconciliation. Since the eavesdropper holds the purification of the state $\rho_{ABE}$, the Von Neumann entropy can be expressed as 
\begin{equation}
\begin{aligned}
S(E)&=S(AB),\\
S(E|x_{B})&=S(A|x_{B})\\
\end{aligned}
\end{equation}
which can be respectively calculated through the covariance matrix $\Gamma_{AB}$, and on the other hand, the conditioned Von Neumann entropy $S_{A|x_{B}}$ through the conditioned entropy $\Gamma_{A|x_{B}}$. More explicitly,
\begin{equation}
\chi_{BE}=G\left(\frac{\lambda_{1}-1}{2}\right)+G\left(\frac{\lambda_{2}-1}{2}\right)-G\left(\frac{\lambda_{c}-1}{2}\right)
\end{equation}
where the function $g(x)$ is defined as
\begin{equation}
G(x)=(x+1)\log(x+1)-x\log x
\end{equation}
and $\lambda_{1,2}$ are symplectic eigenvalues of $\Gamma_{AB}$ and $\lambda_{c}$ is symplectic eigenvalue of $\Gamma_{A|x_{B}}$.

In the description of entanglement-based scheme, the covariance matrix of a two-mode squeezed vacuum state is
\begin{equation}
\gamma=\left[
\begin{array}{ccc}
V\mathbb{I}_{2} & \sqrt{V^{2}-1}\mathbf{\sigma}_{z}\\
\sqrt{V^{2}-1}\mathbf{\sigma}_{z} & V\mathbb{I}_{2}\\
\end{array}\right]
\end{equation}
The variance V is equal to $\sqrt{V_{M}+1}$ in prepare-and-measure scheme with the modulation variance of $V_{M}$. According to \cite{unidimension}, the covariance for unidimensional protocol is built by a squeeze operation on one of its mode, for example, here $S_{1}$, with a squeezing parameter of $r=-\log\sqrt V$, then the covariance matrix reads:
\begin{equation}
\begin{split}
\gamma_{AB}=& S \gamma S^{T}\\
=&\left[
\begin{matrix}
\sqrt{V_{M}+1} & 0 & V_{M}\sqrt{V_{M}+1} & 0\\
0 & \sqrt{V_{M}+1} & 0 & -\frac{V_{M}}{\sqrt{V_{M}+1}}\\
V_{M}\sqrt{V_{M}+1} & 0 & V_{M}+1 & 0\\
0 & -\frac{V_{M}}{\sqrt{V_{M}+1}} & 0 & 1\\
\end{matrix}\right]
\end{split}
\end{equation}
where the squeezing operator S is
\begin{equation}
S=\left[
\begin{array}{ccc}
(V_{M}+1)^{1/4} & 0\\
0 & (V_{M}+1)^{-1/4}\\
\end{array}
\right]
\end{equation}
Now assume that the channel transmittance and noise in X(or equivalently, $S_{1}$) is $\eta_{x}, \epsilon_{x}$, respectively. After transmission through the noisy channel, the covariance becomes
 \begin{equation}
 \label{gammaAB1}
 \gamma^{'}_{AB_{1}} =\left[
 \begin{matrix}
 \sqrt{V_{M}+1} & 0 & \sqrt{\eta_{x}V_{M}}({V_{M}+1})^{\frac{1}{4}} & 0\\
 0 & \sqrt{V_{M}+1} & 0 & C_{p}\\
 \sqrt{\eta_{x}V_{M}}({V_{M}+1})^{\frac{1}{4}} & 0 & 1+\eta_{x}(V_{M}+\epsilon_{x}) & 0\\
 0 & C_{p} & 0 & V_{P1}\\
 \end{matrix}\right]
 \end{equation}
Since the P (or $S_{2}$) quadrature is unmodulated, its variance $V_{p}$ and the correlation between X (or $S_{1}$) and P (or $S_{2}$), $C_{p}$, remains unknown to all communication parties, including eavesdropper. $V_{p}$ is monitored at Bob's side. Considering the realized model of balanced homodyne detectors(BHD), it has a non-unity detection efficient $\eta_{e}$, and electronic noise $V_{e}$ in shot-noise unit. It is modeled as an ideal BHD, followed by a PBS of a transmittance efficient $\eta_{e}$. A thermal state of a noise variance of $1+\frac{V_{e}}{1-\eta_{e}}$ injects from one of the port of PBS. In this case, the covariance of Alice and Bob is \cite{unidimension exp}
\begin{equation}
\gamma^{'}_{AB} =\left[
 \begin{matrix}
\sqrt{V_{M}+1} & 0 & \sqrt{\eta_{e}\eta_{x}V_{M}}({V_{M}+1})^{\frac{1}{4}} & 0\\
0 & \sqrt{V_{M}+1} & 0 & \sqrt{\eta_{e}}C_{p}\\
\sqrt{\eta_{e}\eta_{x}V_{M}}({V_{M}+1})^{\frac{1}{4}} & 0 & 1+\eta_{e}\eta_{x}(V_{M}+\epsilon_{x})+V_{e} & 0\\
0 & \sqrt{\eta_{e}}C_{p} & 0 & \eta_{e}V_{P1}+(1-\eta_{e})+V_{e}\\
\end{matrix}\right]
\end{equation}
Since the unknown parameters $C_{p}$ and $V_{p}$ must be physical, the Heisenberg Uncertainty principle gives their bound
\begin{equation}
\gamma_{AB}+i\Omega\ge 0
\end{equation}
where
\begin{equation}
\Omega=\left[
\begin{matrix}
0 & 1 & 0 & 0\\
-1 & 0 & 0 & 0\\
0 & 0 & 0 & 1\\
0 & 0 & -1 & 0\\
\end{matrix}\right]
\end{equation}
In the assumption of reverse reconciliation \cite{reverse rec}, Alice guesses the measurement of Bob's, so she holds the conditioned covariance matrix:
\begin{equation}
\gamma_{A|x_{B}}=\left[
\begin{matrix}
\frac{\sqrt{V_{M}+1}(1+\eta_{x}\epsilon_{x})}{1+\eta_{x}(V_{M}+\epsilon_{x})} & 0\\
0 & \sqrt{V_{M}+1}\\
\end{matrix}
\right]
\end{equation}

\section{experiment setup}
\begin{figure}
	\begin{center}
		\includegraphics[width=\linewidth]{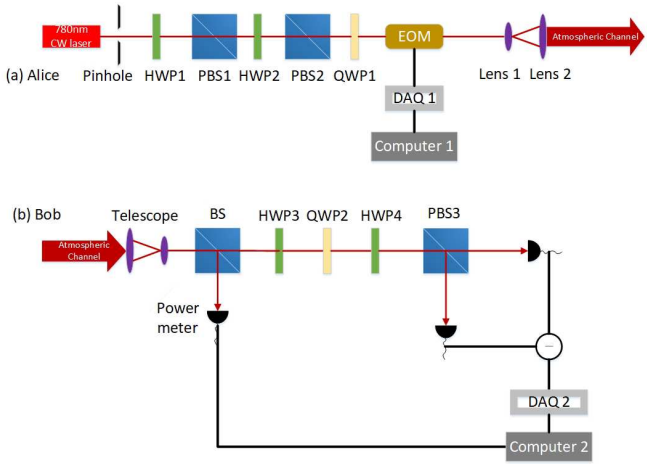}
		\caption{Free-space CV QKD experiment setup for unidimensional protocol. }
		\label{setup}
	\end{center}
\end{figure}

The experiment setup for the unidimensional CV QKD system in free-space is shown in \ref{setup}.

The laser(type) centered at 786nm, about 1 nm of full width at half maximum, is fiber pig-tailed and coupled to free-space in Gaussian mode. The output intensity is 15mW, and passes through two combinations of a PBS(Thorlabs PBS252, extinction ratio$\geq33dB$) and a half-wave plate whose axis is nearly perpendicular to the PBS axis. Since the axises of these two PBSs are aligned at the same direction, and the half-wave plate between them rotates the polarization of light nearly 90 degree, the output light intensity attenuates to $100\mu W$ level, equally the magnitude of orders about $1\times10^{14}$ photons per second. The quater-wave plate rotates the linear polarization light to a certain elliptic polarization, so as it becomes a circle polarization mode after the EOM since the ordinary and extraordinary axises are probably not at the same direction as PBS. Then, the polarization state is modulated in the EOM(Thorlabs, EO-AM C1, wavelength 600-900nm), whose modulation bandwidth is 100 MHz \cite{EOM datasheet}. Since the intensity of signal light is far weaker than local oscillator light (3 magnitude of orders), the voltage is controlled by computer, with Gaussian distributed random intensity whose variance is $\Sigma^{2}$ as mentioned in Section \ref{Polarization encode principle}, which is about 165 times to the shot-noise, at the modulation frequency of 10kHz, which is limited by the acquisition bandwidth of DAQ module(NI PCIe-6363, maximum acquisition rate is 2MHz). The width of modulation signal pulses is 10$\mu s$, which is much longer than the time difference of  light distance between the two arms of BHD (about$10^{-11}s$).

The output signal with the local oscillator beam in spatial mode is focus within 2.1mm of $1/e^{2}$ diameter, and the full divergence angle of $4.5\times10^{-4}$, enter a Galileo beam expander(Thorlabs, GBE10-B, expansion is 10x) so that the transmitted full divergence is $4.5\times10^{-5}$. After propagated through a 460m free-space channel, the beam diameter at the receive aperture is about 4 cm.

At the receiver side, a 4-inch reflection mirror is to adjust beam direction. Two convex lenses whose diameter and focus are 4 inches, 20cm and 2 inches, 6cm reduce the beam diameter to about 1cm. HWP3 is used to calibrate the polarization direction such that the receiving local oscillator is polarized along vertical mode. QWP2 changes linear polarized light to circle so that the photon numbers of two ports of PBS corresponds to left and right circle polarized modes, respectively. Then, another EOM placed between QWP2 and PBS, acts as a basis switcher, to monitor the variance of $S_{2}$ quadrature. It is modulated randomly with the voltage of 0 or $V_{\pi}$. When applied voltage is 0, the measured phonon number is X quadrature, and when $V_{\pi}$, P quadrature. Two congruent convex lens focus the beam into photon diodes (Hamamatsu, S3883, photon-sensitivity 0.58A/W at 780nm, equal to detective efficient 0.872 for each individual diode \cite{hamamatsu datasheet}). Another DAQ module is used to acquisit the output of the difference voltage of two diodes for every pulse, at the sampling rate of 1 MHz. Thus, for each pulse it takes 100 samples, of which 10 for each signal pulse since the duty cycle is set to be 0.1. The average voltage of these 10 sample is raw key value for Bob. A consecutive 5 large pulse (10V) marks the start of each communication, in other words, the pulses followed the start pulses are as the distributed keys.

\section{experiment result}
First we record the beam spot behaviors caused by wandering and vibrating of the buildings. At receiver's side, a CCD camera beam profiler(Thorlabs, BC106N-VIS/M) at the lens focus records the profile and jitter of beam, as shown in Fig. \ref{profile}.
\label{profile}
\begin{figure}
\begin{center}
\includegraphics[width=\linewidth]{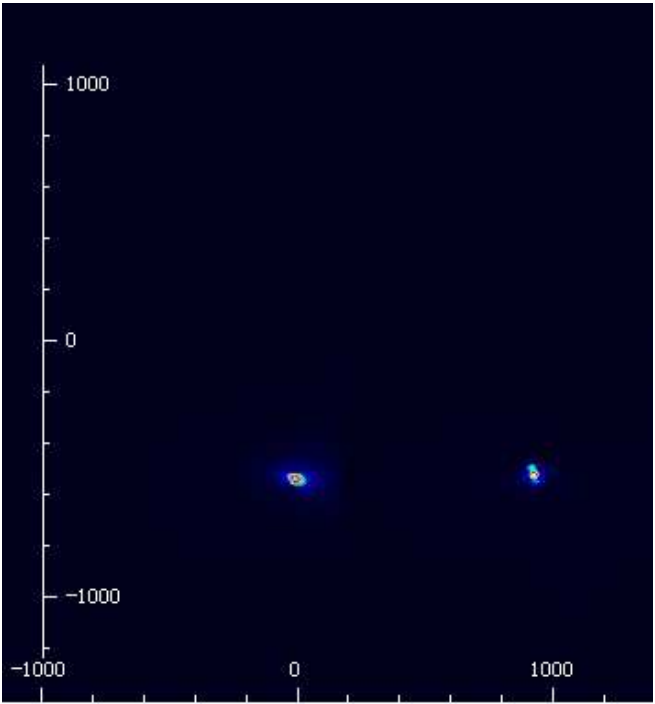}
\caption{Beam profiles recorded by CCD camera after two output ports of PBS at Bob's side at lens focus. Left: vertical polarized. Right: horizontal polarized. For convenience, the two beam profiles are seen in one figure. The horizontal and vertical axises are position of beam, in unit of $\mu m$. The diameter of two beams are not exactly the same because the CCD deviates from lens focus at the two directions. The profiles are Gaussian in both x and y direction but shaped as ellipse since beam is not vertical to lens.}
\end{center}
\end{figure}
The sensitivity area of photo-diodes is about 1.5mm, much larger than beam diameter so can collective all light intensity. However, the intensities are still fluctuate due to atmospheric turbulence.

The sender's side is placed at 9th floor of a building while receiver's side is at 16th floor of another building. Since the height of both buildings is high, their vibration is not negligible. The jitter of beam spot and trace of spot center are shown in Fig. \ref{jitter} and Fig. \ref{trace}, respectively.
\label{jitter}
\begin{figure}
	\begin{center}
		\includegraphics[width=\linewidth]{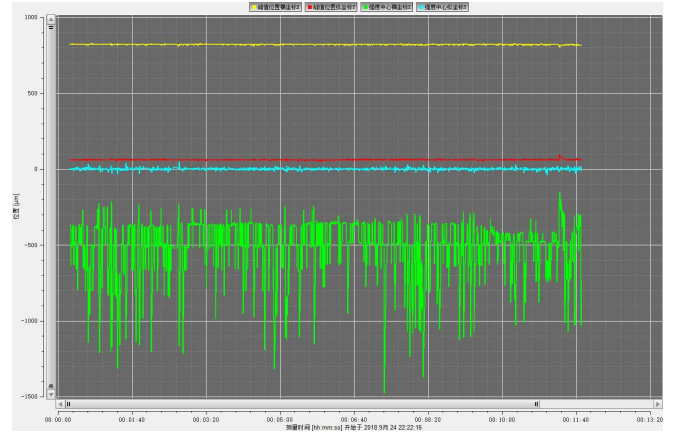}
		\caption{The center of x and y as a function of time of beam at the transmission port of PBS. Measure time is 10 minutes. Red: peak position in y direction. Yellow: peak position in x direction. Blue: position of intensity center in y direction. Green: position of intensity center in x direction. The maximum jitter in vertical direction is about 100 $\mu m$(blue line), while 1000 $\mu m$(green line) in x direction.}
	\end{center}
\end{figure}
The jitter in vertical direction is mainly caused by beam wandering while the buildings vibration in horizontal direction. The frequency of building vibration is much lower than beam wandering due to atmospheric turbulence.
 \label{trace}
 \begin{figure}
 	\begin{center}
 		\includegraphics[width=\linewidth]{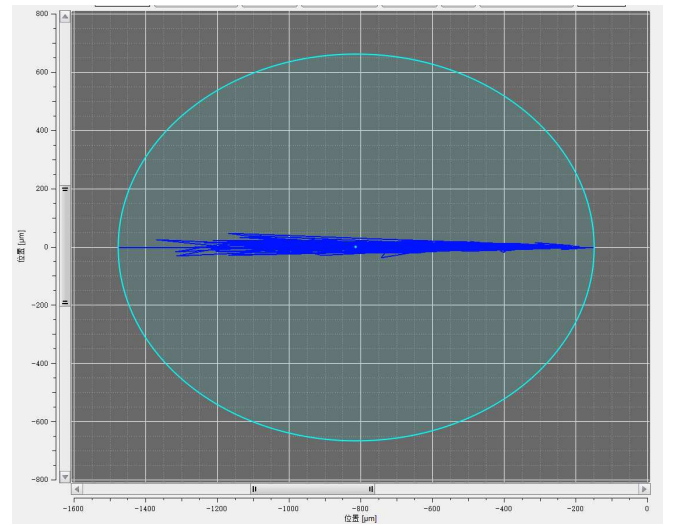}
 		\caption{Positions of beam spot center, unit in $\mu m$. Jitter in horizontal direction is larger than vertical since the amplitude of building vibration is much larger than beam wandering.}
 	\end{center}
 \end{figure}

Although beam spot always varies, it is at order of magnitude of $100~1000 \mu m$, and the beam diameter focused by lens is about 200$\mu m$ according to Fig.\ref{profile} and Fig.\ref{trace}, therefore the whole intensity cannot always impinges on the sensitivity area of photo-diodes, which is of 1.5mm diameter. However, since the homodyne detector subtract two intensities of transmission and reflection output of PBS, the jitter of differential intensity is suppressed.

Before modulate the signal, the shot-noise and electro noise must be measured by a DAQ. The intensity of lase output from EOM is $100 \mu W$, while $65 \mu W$ input to photo-diodes. When the laser is turned off, the variance of measured data is electro noise $V_{e}N_{0}$, in the unit of  $V^{2}$. Then turn on the light, when the detection is balanced, the variance is $N_{0}(1+V_{e})$. Subtracting two variance gets the shot noise $N_{0}=15.4 mV^{2}$, and thus $V_{e}=0.0219$. Then $5\times10^{5}$ Gaussian distributed (pseudo) random variable, centered at 0, of variance of $1V^{2}$, are generated by computer software, and pulsed by the output of DAQ. Modulation variance $V_{M}=165$ when $\Sigma=1 V$.
 However, smaller $V_{M}$ would be comparable to the leaked light from the local oscillator since the isolation ratio of a single PBS is only about 33 dB. The total transmittance measured with power meter is 0.65, including optical components reflecting loss and channel loss. A random chosen part about $1/5$ of data is used to estimate channel parameters $T$ and $\epsilon$ based on the equations:
\begin{equation}
\begin{aligned}
\tilde{T}&=(\frac{Cov(x,y)}{V_{M}})^{2}\\
\tilde{\epsilon}&=\frac{Var(y)-V_{e}-1}{\eta\tilde{T}}-V_{M}\\
\end{aligned}
\end{equation}
where x, y are the chosen string of data of Alice and Bob's. The excess noise $\tilde{\epsilon}=0.0375$ and $\tilde{T}=0.575$, and the latter is lower than that measured by power meter since the sensitivity area diameter of power meter is much larger than photo-diodes.

Another random $1/5$ part of data is used to monitor the variance of P quadrature, controlled by EOM2. When the applied voltage is $V_{\pi}=284 V$, the EOM2 acts as a half wave plate and rotate the polarization direction by $45\deg$ to measure $\pm45\deg$ modes. The accuracy of applied voltage is 0.1 V, $3.52\times10^{-4}$ of $V_{\pi}$, precise enough to suppress the modulated signal. Therefore, $V_{P1}$ of Eq. \ref{gammaAB1} is 1.00.

With all parameters achieved above, the security key rate can be evaluated. At distance of 460m atmospheric environment, secret key rate is 0.0254 bit per pulse at a typical reconciliation efficiency of 0.95, corresponding to 0.152 kbps, while secret key rate is 0.23 bit per pulse in laboratory environment at the same modulation voltage, electro noise but lower local oscillator intensity(30$\mu W$). The expected secret key rate of unidimensional versus GG02 protocols at different total loss with experiment measured parameter $V_{M}, \epsilon, V_{e} and V_{P1}$ is shown in Fig. \ref{comparison}.
\label{comparison}
\begin{figure}
	\begin{center}
		\includegraphics[width=\linewidth]{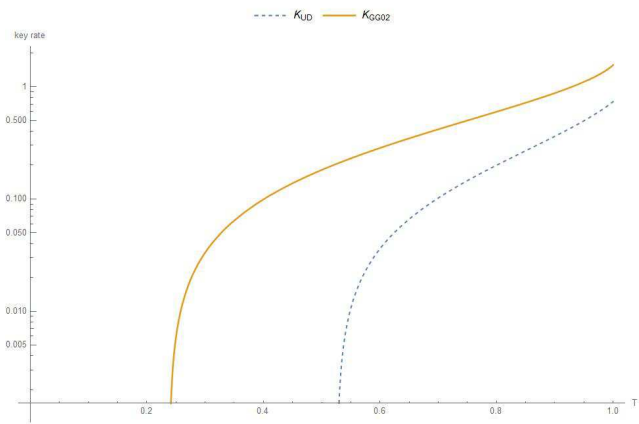}
		\caption{Key rate at different channel transmittance. Solid line: GG02; dashed line: unidimensional protocol. Main parameters: $V_{M}=165, \epsilon=0.0375, V_{P1}=1.0, V_{e}=0.0219, \eta=0.872$}
	\end{center}
\end{figure}
At low channel loss, unidimensional performs close to GG02 protocol, but when transmittance is less than about 0.6, it is a magnitude of order lower than GG02. 

In our experiment, the main factors to restrict secret key rate are the sample rate of DAQs, and the polarization fluctuation of laser. As mentioned above, sample rate of DAQs is up to 1 MHz, and modulation frequency is even lower, far less than the response bandwidth of EOM, 100 MHz. Besides, the memory and CPU of computer at Bob's side are not able to process too many key data (over $10^{7}$), the key rate considering finite-size region is expected to be further lower than the asymptotic limit\cite{ud finite key}. At last, the fiber pig-tailed laser coupled to free-space,  may cause the polarization direction changing in the fiber and be unstable, lead to polarization noise and thus intensity fluctuation of $0.1\%$. The further improvement will be made to extend the secure distance.

\section{conclusions}
To conclude, we accomplished free-space unidimensional CV QKD experiment in a real urban environment through the atmospheric channel of 460m. In such a condition, the variance of the unmodulated quadrature, $S_{2}$ barely remains unchanged. With the correlation of two quadrature unknown, the pessimistic raw key rate against collective attacks reaches 0.0254 bit per pulse, at the modulation repetition of 10 kHz.

\end{document}